\documentclass[preprint,showpacs,preprintnumbers,amsmath,amssymb]{revtex4}

 \usepackage{epsfig}
\usepackage{amssymb}
\usepackage{graphicx}

\begin{document}
\title{Interaction of Ultracold Antihydrogen  with a Conducting Wall }
\author{A. Yu. Voronin, P. Froelich, B. Zygelman}
\affiliation{ P.N. Lebedev Physical Institute, 53 Leninsky
prospect, 117924 Moscow, Russia
\\
Department of Quantum Chemistry, Uppsala University, Box 518,
SE-75120 Uppsala, Sweden.}

\begin{abstract}

We investigate the interaction of ultracold antihydrogen with
a conducting surface. Our discussion focuses on the physical regime where the
phenomenon of quantum reflection manifests.
We calculate the reflection probability
as function of incident atom energy.  We
find that, for ground state $\bar{H}$ atoms (with $T<
10^{-5}$ K), the probability of reflection is $R \simeq 1-k b$,
where $k$ is the momentum of the atom and $b = 2174.0$ a.u. is a
constant determined solely by the van der Waals-Casimir tail of the
atom-wall interaction.

We show that quantum reflection, which suppresses the direct contact
of ultra-cold atoms with the surface, allows for the possibility of
confinement and storage of cold antihydrogen atoms. We calculate the
life-time of confinement as a function of antihydrogen energy. We
develop a theory of $\bar{H}$ in a wave-guide and propose its
application to fundamental measurements. In particular, for
measurement of retardation corrections in the long-range component
of the antiatom - wall potential.
 We demonstrate, for $\bar{H}$ falling in the gravitational field of Earth onto a conducting surface,
 the existence of quantized  $\bar{H}$ states.
 We calculate  that the lifetime of  ultracold $\bar{H}$
in its lowest gravitational state and obtain  $\tau=(Mg b/2\hbar)^{-1}\simeq
0.1$ s, where $Mg$ is a gravitational force acting on the antiatom.
We propose that measurement of this lifetime may provide a new
 test for the gravitational properties of antimatter.

\end{abstract}

\maketitle

\section{Introduction}

The phenomenon of quantum reflection in the ultra-cold atomic
collisions has attracted lots  of attention during the last decade.
The propensity of an ultra cold atom, under the influence of an
\emph{attractive} atom-wall potential, to undergo total reflection
has been predicted on
quantum-mechanical grounds quite some time ago \cite{LJ}. However,  the
observation of this counter-intuitive effect has
only recently been observed
in the laboratory \cite{yu93}, \cite{ExCh1}. Because quantum reflection prevents
atoms from reaching the wall,
  it suppresses inelastic atom-surface reactions
and thus, it can be exploited as an important tool for manipulating cold atoms
\cite{ExCh0}. Another reason for interest in quantum reflection
is that it occurs at large atom-wall distances where
retardation effects are important. Retardation
leads to the Casimir effect
\cite{CP,DLP,Bord}, where the
 van der Waals $1/z^3$ atom-surface potential melds into a $1/z^4$ power law potential
 at large distances.
 Quantum reflection is
sensitive to the long-range component of the atom-surface potential
\cite{FJM} and  may therefore  allow new tests for the
predictions of QED \cite{ExCh1,ExCh2,ExCh3,ExCh4}.

  Recent success in the production of the
cold antihydrogen (further referred to as $\bar{H}$) \cite{AH1,AH2}
has spurred renewed interest in this field. The ultimate goal
of the antihydrogen project, initiated at CERN, is to enable
accurate tests of CPT and gravitational properties of antimatter. Such a capability
requires the availability of ultra-cold antiatoms. Quantum
reflection may provide a  new  tool that hastens the realization of laboratory measurements.
  We will show that an
interesting manifestation of quantum reflection, for antiatoms,
is the existence of long-lived states of ultracold $\bar{H}$ in
material cavities. We introduce a theory of wave-guides for anti-atoms
and discuss their possible application in laboratory studies.
 Measurement of anti-atom life-times in a cavity may provide valuable
information on the properties of the long range interaction between the
antiatom and the surface.
The measurement of the gravitational force acting on an
antiatom poses as an intriguing possibility. An essential property of antiatom-surface interaction is
annihilation of $\bar{H}$ on the surface.  Anti-atoms that are not reflected are
 lost to annihilation and $\bar{H}$-wall annihilation events
can be detected  to measure the quantum reflection
probability.

In Sec. II we discuss quantum reflection of $\bar{H}$
impinging on a perfectly conducting surface at normal incidence and
present a calculation for the reflection probability. The latter is
given as a function of incident energy, including the zero energy
limit characterized by the \emph{complex} scattering length. In
section III we study the probability for quantum
reflection as function of the antiatom - wall distance. In Sec. IV
we apply our theory  and calculate  the lifetimes of
$\bar{H}$ enclosed in a  material cavity. In Sec. V we discuss the
passage of $\bar{H}$ through wave-guides.  In Sec. VI we discuss
the possibility for application of quantum reflection in measurements of the
gravitational properties of an antiatom. Finally, in Sec. VII we
study the loosely bound states of atoms confined in the atom-wall
potential. We show that such states strongly affect quantum
reflection when absorption by the wall is weak.
We discuss the physical conditions corresponding to the limiting
  cases of weak and strong absorption (annihilation) on the surface and
  show that, in
the latter case, the (anti)atom-wall states are destroyed.

\section{Quantum reflection and absorption}

\label{qr}
  Quantum reflection from an \emph{attractive} (with a sufficiently large derivative)
potential is known to  manifest
  \cite{frie04a} at low energies. In our discussion we are mainly interested
  in those cases where the probability of quantum reflection is
  close to 1. The qualitative parameter associated with the quantum motion of
the particle with mass $m$ and energy $E$ in the
  potential $V(z)$ is the local de
Broglie wavelength $\lambda_{B}(z)=2\pi\hbar/\sqrt{2m(E-V(z))}$. One
expects quantum behavior when:
\begin{equation} \label {dB}
\frac{\partial \lambda_{B}(z)}{\partial z}\geq 1.
 \end{equation}
For a homogeneous potential, $-C_s/z^s$ with $s>2$ in the limit of
zero energies condition (\ref{dB}) is fulfilled for distances $z$
such that
\begin{eqnarray}
  z_c&\leq& z \ll z_f \label{range} \label{zc} \\
  z_c&\equiv&(2\sqrt{2m C_{s}}/s)^{2/(s-2)} \\
  z_f&\equiv &1/\sqrt{2mE}.
\end{eqnarray}

Condition (\ref{range}) defines the domain where the reflected wave
is generated and is applicable if the energy of the incident atoms
satisfies the condition
  \begin{equation} \label{Ec}
E\ll E_c=1/(2m z_c^{2}).
  \end{equation}
A detailed discussion of the ultra-cold atom quantum reflection can be found in refs.
\cite{FJM,juri04,frie04}.
The $\bar{\rm H}$ -- wall interaction is described by the potential,
hereafter denoted $V(z)$. It is {\it not} homogeneous and differs
from that of the H - wall interaction. In particular it is purely
attractive even at very short distances $z \leq z_s\simeq 1$ a.u.
and is strongly absorptive at the origin due to the likely-hood of annihilation.
However at large $\bar{\rm H}$ -- wall separations $z\gg z_s$ it
is dominated by induced dipole-dipole terms in the
atom-wall interaction, similar to the case of the hydrogen - wall system.
The potential for ${\rm H}$  impinging on a perfect
conducting wall is known for large $z$ \cite{CP} and will
be hereafter denoted as $V_{CP}(z)$;  in  our treatment we use
$V_{CP}(z)$ calculated in \cite{DMB}. At distances $z_s \ll z \ll
\lambda_{\omega}$ the potential has the  van der Waals form
\begin{equation}\label{vdW}
V_{CP}(z)\simeq -C_3/z^3
\end{equation}
with
  $C_3= \frac{1}{4\pi}\int_{0}^{\infty} \alpha_{d}(i\omega) d\omega=0.25 \mbox{
  a.u.}^2$.
Here $\alpha_{d}(i\omega)$ is the dynamic dipole polarizability of
the (anti)hydrogen atom, expressed as a function of imaginary frequency
$i\omega$, and $\lambda_{\omega}$ is the effective wavelength
that gives the main contribution to $\alpha_{d}(i\omega)$.
  For the distances $z\gg \lambda_{\omega}$
retardation effects are important and the potential is given by
\begin{equation}\label{Cas}
V_{CP}(z)\simeq -C_4/z^4
\end{equation}
where $C_4=\frac{3}{8\pi}\frac{\alpha_d(0)}{\alpha}=73.62 \mbox{
a.u.}^3$,
  $\alpha_d(0)=9/2$ a.u.$^3$ is the ground
state static dipole polarizability of the antihydrogen, and
$\alpha=1/137.04$ is the fine-structure constant. We note
that the above expressions for $C_3$ and $C_4$ are valid in
the limit of perfect conducting surface \cite{CP}, and can be specified
for a realistic metal or dielectric surface at finite temperature
\cite{BKM}.
At short distances $z\simeq z_s$ the interaction between $\bar{H}$
and the wall differs from that of $H$. In particular it includes
an inelastic component, corresponding to the process of
capture of $\bar{p}$ and $\bar{e}$ in the medium of the wall followed
by subsequent annihilation. We take annihilation into account
by imposing full absorption on the wall
implemented in the manner described below.

The Schr\"{o}dinger equation that governs the $\bar{H}$-wall
scattering is
\begin{equation} \label{Schr}
\left[ -\frac{\partial ^{2}}{2m\partial z^{2}}+V(z)-E\right] \Phi
(z)=0
\end{equation}
At distances $z_s \ll z \ll \lambda_{\omega}$ the potential
has form (\ref{vdW}) and the solution of Schr\"{o}dinger equation is
\begin{eqnarray}
\Phi (z) &\sim &\sqrt{z}\left(H_{1}^{(1)}(\rho)+\exp(2i\delta_s)
H_{1}^{(2)}(\rho)\right) \label{Phi} \\
\rho &=&2\sqrt{2mC_3/z}
\end{eqnarray}
where $H_{1}^{(1)}(\rho)$ and $H_{1 }^{(2)}(\rho)$ are the Hankel
functions \cite{W} of order $1 $, and
$\delta_s=\delta_1+i\delta_2$ is a complex phase-shift produced by
the short range part of the interaction. The imaginary part of
$\delta_s$  is due to annihilation of $\bar{H}$ on the surface. We
 used here the analytical form of the zero-energy solution in
homogeneous potential $-C_3/z^3$ \cite{MM} and neglected the energy of
the incident $\bar{H}$ since it is small compared to the potential
$V_{CP}(z)$ at $z \simeq z_s$.
  In our treatment we are interested in low energies where
the above requirements are always satisfied.

The condition of full absorption, which we apply, is
\begin{equation}\label{fullabs}
\delta_2\gg 1.
\end{equation}
This condition selects the "incoming wave" solution of the
Schr\"{o}dinger equation and suppresses the "reflected wave"
\begin{equation}
\label{IW} \Phi (z) \sim \sqrt{z} H_{1}^{(1)}(\rho)\mbox{ , } z_s
\ll z \ll \lambda_{\omega}.
\end{equation}
One can check that in the limit of small $z$ the above solution
coincides with the semiclassical "incoming wave"
\begin{equation}\label{IWs}
\sqrt{z} H_{1}^{(1)}(2\sqrt{2MC_3/z})\rightarrow
\frac{1}{\sqrt{p(z)}}\exp(-i\int p(x)dx)
\end{equation}
where $p(z)=\sqrt{2mC_3/z^3}$ is the local classical momentum.

The full absorption condition $\delta_2\gg 1$ leads to
 insensitivity of solution
(\ref{IW}) in  details of the short-range
interaction since it is independent of the short-range phase shift $\delta_s$.
Thus, for strong annihilation on the
surface  the outgoing flux of $\bar{H}$ is determined solely by the asymptotic
properties of the potential.

We impose (\ref{IW}) as a boundary condition in solution of the
Schr\"{o}dinger equation, which  now does not depend on the
short-range physics. The large $z$ asymptotic form of such a
solution is
\begin{equation}
\Phi(z\rightarrow \infty) \sim \exp(-i k z)-S\exp(i kz)
\label{outerbc}
\end{equation}
where $k=\sqrt{2mE}$ is the momentum of incident $\bar{H}$ atoms and
$S$ is the diagonal element of the $S-$matrix that describes elastic
scattering. $S$ defines the flux reflected by the asymptotic
potential  and the reflection coefficient $R = |S|^2$
%
%
has been obtained by numerical integration of eq. (\ref{Schr}) with
eq. (\ref{IW}) as a boundary condition. The resulting reflection
coefficient is presented in Table \ref{Table1} as a function of the
incident energy. It is seen from the last column that the reflection
becomes considerable ($> 50\,\%$) at $E\simeq 10^{-11}$ a.u.
($T\simeq 2.7 \cdot 10^{-6}$ K ) and reaches $99\,\%$ at $E\simeq
10^{-14}$ a.u. ($T\simeq 2.7 \cdot 10^{-9}$ K).
  The results underscore that, in spite of
strong atom-wall attraction, slow (anti)atoms are reflected by
and only a small
fraction of the incoming flux reaches the wall.

The reflection coefficient in the low energy limit
is conveniently expressed in terms of the
energy independent constant $  a = \lim_{ k \rightarrow 0}
\frac{1-S}{2ik} $, known as the scattering length. In the low energy limit
the amplitude of the reflected wave $S$ can be expanded in terms of $k$
\begin{equation}
S(k) \simeq 1 - 2ika. \label{slapp}
\end{equation}
Due to  annihilation on the wall $|S|<1$, the scattering
length $a$ acquires
  a negative imaginary part. The flux reflected in the region
  where the long range tail of the interaction potential dominates
takes the form frequently used in the literature on quantum
scattering \cite{FJM}
  \begin{equation}
|S|^2\simeq 1-4k|\mathop{\rm Im} a|=1-kb
  \end{equation}
  where we have introduced the constant $b=4|\mathop{\rm Im} a|$.
  The flux, absorbed (annihilated) on the wall is
  \begin{equation}
 P(k)=1-|S|^2\simeq kb. \label{P}
  \end{equation}
The annihilation probability $P$ becomes small when $kb\ll 1$, whereas
the reflection probability tends to 1.
We have calculated the scattering length for the potential
$V_{CP}(z)$ with full absorption boundary condition,
\begin{eqnarray}\label{aCP}
a_{CP}&=&-52.4 -i543.5 \mbox{ a.u.}, \\
\mathop{\rm Re}a_{CP}/\mathop{\rm Im}a_{CP}&=&0.09 \label{ReImCP},
\\ b&=& 2174.0 \mbox{ a.u.}.
\end{eqnarray}
The annihilation probability $P$ for potential $V_{CP}$ is expressed in
terms of the scattering length and is tabulated in Table \ref{Table1}
along with the exact values. One notices from column 4 that the
scattering length approximation (eq. (\ref{P}) becomes valid for
$E<10^{-11}$ a.u. i.e. when $kb \ll 1$, but the validity of the
non-perturbative approximation (given in column 3) extends to higher
temperatures.
\begin{table}
\centering
\begin{tabular}{|c|l|l|l|l|}
   \hline
   $\log(E/a.u.)$ & $P$ & $1-\exp(-kb)$ & $kb$ & $R$ \\
   \hline
   -9 & 0.95 & 0.99 &4.16& 0.05 \\
   -10 & 0.69 & 0.74 & 1.32& 0.31\\
   -11 & 0.33 & 0.34 &0.42&0.67 \\
   -12 & 0.12 & 0.13 & 0.13&0.88\\
   -13 & 0.04 & 0.04 &0.04 &0.96\\
   -14 & 0.013 & 0.013&0.013&0.987\\
   -15 & 0.0042 & 0.0042&0.0042&0.9958 \\
   -16 & 0.0013 & 0.0013&0.0013&0.9987 \\
   -17 & 0.00042 & 0.00042 &0.00042&0.99958 \\
   -18 & 0.00013 & 0.00013 &0.00013&0.99987\\ \hline
\end{tabular}
\caption{The annihilation $(P)$ and reflection $(R)$ probabilities
for the ultra-cold antihydrogen impinging on the material wall.
Column 3: non-perturbative approximation to $P$. Column 4:
scattering length approximation to $P$.} \label{Table1}
\end{table}

It is interesting to compare this value for the scattering
length with the corresponding value for the purely homogeneous
potential $-C_4/z^4$ with the full absorption at the origin.
In the latter case \cite{AV}
\begin{equation}
a^{abs}_s=\exp (-i\pi /(s-2))\left( \frac{\sqrt{2m
C_{s}}}{s-2}\right) ^{2/(s-2)}\Gamma ((s-3)/(s-2))/\Gamma
((s-1)/(s-2)) \label{sclength}
\end{equation}
and for $s=4$ (the Casimir correction)  the scattering length is
purely imaginary and given by
\begin{equation}\label{a4}
a_4=-i\sqrt{2mC_4}
\end{equation}
which for $\bar{H}$ results in
\begin{equation}
a_4(\bar{H})=-i519.9 \mbox{ a.u.}.
\end{equation}
We note that the imaginary part of the scattering length $a_{CP}$
is rather close to the value obtained for a purely homogeneous
$-C_4/z^4$ potential and suggest that the main contribution to the
penetration probability arises from the asymptotic Casimir tail
(\ref{Cas}). The nonzero real part of the scattering length
is the contribution from  distances where the potential
changes from the van der Waals (\ref{vdW}) to the Casimir
limit (\ref{Cas}). In fact, one can see from (\ref{sclength}) that
the real part of the scattering length for the purely homogeneous
potential with $s=4$ (with absorptive core) is exactly zero, while
for $s<4$ it is negative.

We investigate the influence of the inner van der Waals part of the
potential in the zero energy limit. We model the exact potential
$V_{CP}(z)$ by the analytically solvable potential \cite{FJM}

\begin{eqnarray}\label{Vm}
V_m(z)&=&-\frac{C_4}{z^3(z+l)}
\end{eqnarray}

with $l\equiv C_4/C_3$. Such a potential has correct asymptotic
behavior in the limit of big ($z\gg l$) and small ($z\ll l$)
distances. The zero energy wave-function for such a potential is

\begin{equation}
\label{Phim} \Phi_m(z)=\sqrt{z(z+l)}\left(H_1^{(1)}
(2\sqrt{2mC_3(1/z+C_3/C_4)})-e^{2i\delta_s}H_1^{(2)}
(2\sqrt{2mC_3(1/z+C_3/C_4)})\right).
\end{equation}

To obtain the scattering length we examine this expression in the
limit $z\gg l$  to first order in terms proportional to $l/z$

\begin{eqnarray}
\label{PhimAs1} \Phi_m(z)&\sim& z(1+l/(2z))\left(H_1^{(1)}(\xi
(1+\frac{l}{2z}))-e^{2i\delta_s}H_1^{(2)}(\xi
(1+\frac{l}{2z}))\right)
\end{eqnarray}

where $\xi=2C_3\sqrt{2m/C_4}$. Employing a Taylor expansion and
collecting terms proportional to $z$, the wave function becomes

\begin{equation}
\label{PhimAs2} \Phi_m(z)\sim
\frac{l}{2}\left(1+\xi\frac{H_1^{'(1)}(\xi)-e^{2i\delta_s}H_1^{'(2)}(\xi)}
{H_1^{(1)}(\xi)-e^{2i\delta_s}H_1^{(2)}(\xi)}\right)+z.
\end{equation}

Now we use the fact that the asymptotic expression of the
wave-function can be written in terms of the scattering length as

\begin{equation}
\Phi_m(z\gg l)\sim 1-z/a_m \label{asymptform}
\end{equation}

Comparing the last two equations we extract the
scattering length from eq. (\ref{PhimAs2})

\begin{equation}\label{amex}
a_m=-\frac{l}{2}\left(1+\xi\frac{H_1^{'(1)}(\xi)-e^{2i\delta_s}H_1^{'(2)}
(\xi)}{H_1^{(1)}(\xi)-e^{2i\delta_s}H_1^{(2)}(\xi)}\right).
\end{equation}

  The full absorption boundary condition ($\delta_2\gg1$,
$e^{2i\delta_s}\rightarrow 0$)  leads to, in analogy with the results of (\ref{IW}),
  cancellation of all
terms that include the  solution $H_1^{(2)}(\xi)$
\begin{equation}\label{am}
a_m
=-\frac{l}{2}\left(1+\xi\frac{H_1^{'(1)}(\xi)}{H_1^{(1)}(\xi)}\right)
\end{equation}
The ratio $\mathop{\rm Re}a_m/\mathop{\rm Im}a_m$ is given by
\begin{equation}
\mathop{\rm Re}a_m/\mathop{\rm Im}a_m =
\frac{\pi}{2}\left(J_1^2(\xi)+Y_1^2(\xi)+\xi(J'_1(\xi)J_1(\xi)+
Y'_1(\xi)Y_1(\xi))\right)
\end{equation}
where $J_1(z)$ and $Y_1(z)$ are Bessel functions \cite{W} of
order $1 $. For $\bar{H}$ we obtain
\begin{eqnarray}
a_m&=&-69.8-i505.6 \mbox{ a.u.}, \\
\mathop{\rm Re}a_m/\mathop{\rm Im}a_m&=&0.14, \\
\xi &=&3.53
\end{eqnarray}
The ratio $\mathop{\rm Re}a_m/\mathop{\rm Im}a_m $ contains
important information on the scale parameter $\xi$  that, in turn,
determines the transition between the van der Waals and
Casimir limits. In the limit of large $\xi$ the model potential
$V_m$ becomes $-C_4/z^4$ whereas the ratio becomes
$\mathop{\rm Re}a_m/\mathop{\rm Im}a_m \simeq 1/\xi\rightarrow 0$ as
it should for the purely homogeneous potential $V=-C_4/z^4$. According
to this treatment, we expect that for the exact potential
$V_{CP}$, the smaller the ratio $\mathop{\rm Re}a_m/\mathop{\rm
Im}a_m $, the more important the contribution from the Casimir tail
 to the scattering amplitude

\section{Probability of quantum reflection as a function of distance}
\label{qr-z}
We now study the relative importance of
different regions in which the van der Waals-Casimir potential dominates
and contributes to the value of the reflected wave. We
represent the wave-function in the following form \cite{Babikov}
\begin{equation}\label{PF1}
\Phi(z)=\frac{A(z)}{p^{1/2}(z)}\left(B(z)\exp(i\int_{z_0}^z p(x)
dx)-\exp(-i\int_{z_0}^z p(x) dx)\right) 
\end{equation}
where $A(z)$ and $B(z)$ are arbitrary functions that are obtained
when we substitute $\Phi(z)$ into the Schr\"{o}dinger equation. Here
$p(z)=\sqrt{2m(E-V_{CP}(z))}$ is the local classical momentum, and
$z_0$ is an arbitrary distance. Identification of a solution of the
Schr\"{o}dinger equation with the incoming or outgoing wave is
unambiguous only in case when it can be represented in the
semiclassical form $\exp(\pm i\int_{z_0}^z p(x) dx)$. If the the WKB
approximation is valid in the entire range of antiatom-wall
distances, the solution everywhere has the form of an incoming wave
and no reflection occurs. However, for certain regions (\ref{zc}) of
antiatom-wall separation, the WKB approximation fails (the so-called
badlands). The solution of the Schr\"{o}dinger equation differs from
the semiclassical one and leads to the appearance of a reflected
wave. Hence the function $B(z)$ in (\ref{PF1}) can be interpreted as
a function that "converts" the semi-classical solution into the
exact quantal solution and thus contains information on the
reflected wave amplitude generated at \emph{each separation
distance} $z$.

We need to put one more condition on $A(z)$ and $B(z)$ in order to define
them uniquely. Following the phase function method
\cite{Babikov,Calog} we require that
\begin{equation}\label{PF2}
\Phi'(z)=iA(z)p^{1/2}(z)\left(B(z)\exp(i\int_{z_0}^z p(x)
dx)+\exp(-i\int_{z_0}^z p(x) dx)\right).
\end{equation}
As we have already mentioned, the function $B(z)$ is an
amplitude for the reflected wave. The relationship between the
asymptotic value $B(\infty)$ and the S-matrix can be easily
established
\begin{equation}
S=B(\infty)\exp(2i(\int_{z_0}^{z\rightarrow\infty} p(x) dx-kz)).
\end{equation}

Substituting eq.
  (\ref{PF1}) and eq. (\ref{PF2}) into the
Schr\"{o}dinger equation we obtain the first order nonlinear
differential equation for $B(x)$:
\begin{equation}\label {RF}
B'(z)=\frac{p'(z)}{2p}\left(B^2(z)\exp(2i\int_{z_0}^z p(x)
dx)-\exp(-2i\int_{z_0}^z p(x) dx)\right).
\end{equation}
As long as in the limit $z\rightarrow z_s$ the wave-function should
have the form given by (\ref{IW}) and (\ref{IWs}) one should choose
the initial condition  as $B(z_s)=0$, and in equation (\ref{RF}) one
should put $z_0=z_s$.

 The function
$|B(z)|^2$ can be interpreted as that "portion" of the reflected
wave that is generated in the domain between at $z_0$ and $z$. (Such
an interpretation is unambiguous when $z$ belongs to the range where
the WKB approximation is valid, and the functions $\exp(\pm
i\int_{z_0}^z p(x) dx)$ can be identified with the reflected or
transmitted waves). Obviously $|B(\infty)|^2$ gives the "full"
reflection probability for the given energy.  In Fig.1 we plot the
function $|B(z)|^2$ for two energies $E=10^{-12}$ a.u. and
$E=10^{-10}$ a.u.


\begin{figure}
  \centering
\includegraphics[width=120mm]{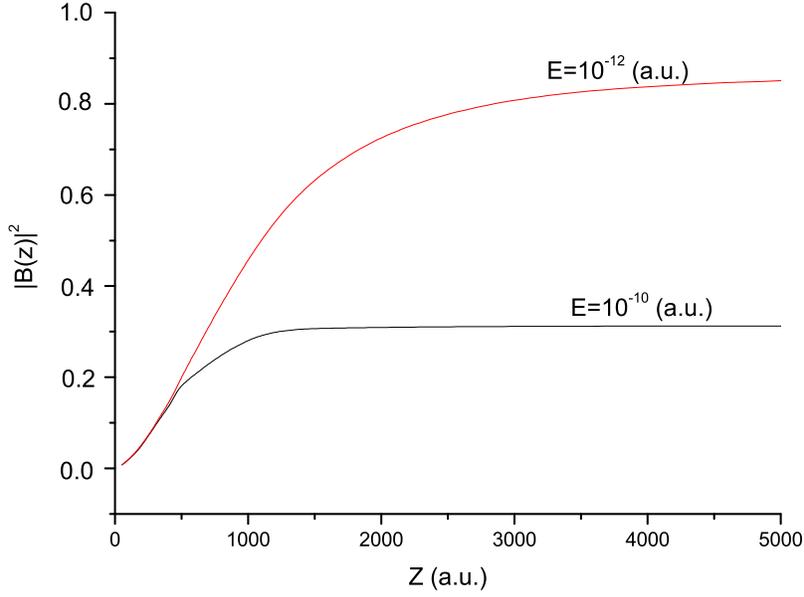} 
\caption{Contribution of different antiatom-wall distances to the
reflection probability. $|B(z)|^2$ expresses the reflection
probability accumulated in the interval between $z_0$ and $z$.
}\label{Fig1}
\end{figure}


As $z\rightarrow 0$  the WKB
approximation becomes valid. Using it
 (and thus reflection from these parts of potential
exponentially decreases) we found that the contribution to the
amplitude of the reflected wave from the distances $z<100$ a.u. is
indeed small.

  In addition, at ultra-low
energies (practically for $E\ll 10^{-10}$ a.u.) the reflected wave
is generated within a wide range of distances. In that
case we cannot define an unambiguous  "reflection
distance". The plot shows that for the energy
$E=10^{-12}$ a.u. the reflection probability is $0.88$ and  this value is
in harmony
with the calculated value presented in Table \ref{Table1}. We note
 that $75$\% of the reflected wave amplitude is
generated in the domain from $500$ a.u. to $5000$ a.u., the region
where the Casimir tail of the potential dominates. At intermediate distances,
from $100$ a.u. up to $500$ a.u., the contribution is $22$\%. However, with
increasing energy the reflecting domain is more
localized and is shifted to shorter distances, whereas the reflection
probability is diminished. According to Fig.1,
for $E=10^{-10}$ a.u. the reflection probability is $0.31$ and
 $92$\% of the reflected wave is generated in the domain from
$100$ a.u. up to $1000$ a.u.  In this case the reflection
probability reaches its full value within a well defined interval
$\Delta z$ around $z_r$, and allows us to define $z_r$, the
"reflection distance". At higher energies, reflection occurs at the
reflection distance ( $\Delta z\ll z_r$ ) and the reflection
probability becomes exponentially small. The reflection coefficient
as well as reflection distance $z_r$ can be estimated using "complex
turning point method" \cite{Pokrovskii1, Pokrovskii2, FJM}.

\begin{figure}

\centering

\includegraphics[width=120mm]{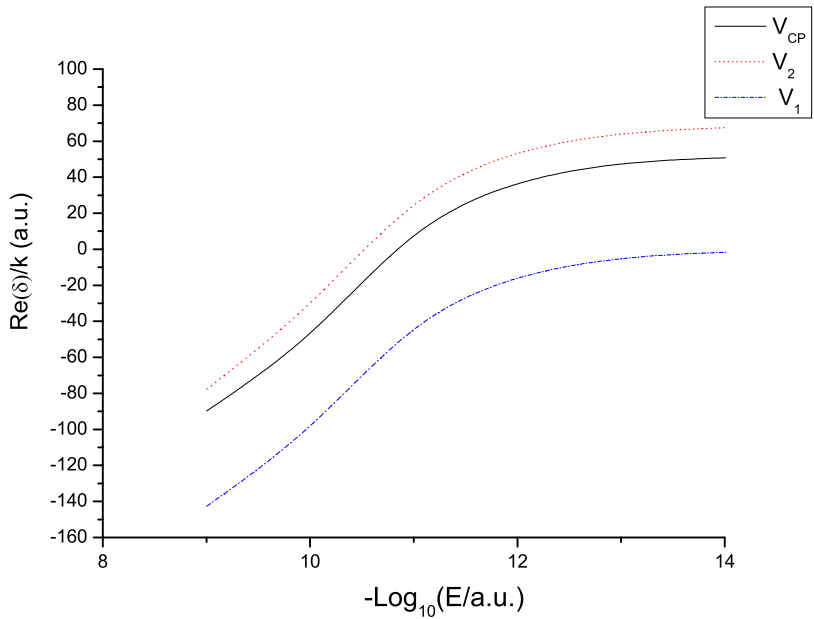}
  \caption{$\mathop{\rm Re}\delta/k$ as a function of energy for
exact interaction $V_{CP}$ (solid line) and for model interactions
  $V_1$ (dashed line) and $V_2$ (dotted line). } \label{Fig2}
\end{figure}

As discussed above, the real part of the scattering length
is more sensitive to the details of the potential at intermediate distances.
More
generally, one can expect that the real part of the phase shift
$\delta(E)$ (defined through the equality
$S=\exp(2i\delta(E)$)) provides additional information on the
antiatom-surface interaction. In Fig. \ref{Fig2} we plot
$\mathop{\rm Re}\delta(E)$ for the exact interaction $V_{CP}(z)$, as
well as for two model interactions $V_1=-C_4/z^4$ and
$V_2=-C_4/(z^4+z^3C_4/C_3)$.


We found that the exact phase shift is negative for $E>10^{-9}$
a.u. and positive for $E<10^{-9}$ a.u. The phase shift produced by
the homogeneous potential $V_1=-C_4/z^4$ is always negative and
tends to $0$ in the limit $E\rightarrow0$. Note that in the limit of
zero energy $\mathop{\rm Re}\delta(E\rightarrow0)/k=-\mathop{\rm
Re}a_{CP}$.

In conclusion,  we emphasize that annihilation of $\bar{H}$ on the
surface allows a description that depends only on the nature of the
van- der Waals-Casimir potential. In particular all inelastic
processes that occur at characteristic distances of few a.u.
(including thermal heating of $\bar{H}$ atoms by phonon exchange
\cite{Phonon1,Phonon2}) do not affect the reflected wave.

In the case of strong absorption the phase and amplitude of the
reflected wave are independent inn details of the short-range
interaction. This allows one to glean important information on
parameters that characterize the van der Waals-Casimir atom-wall
potential.  In the limit of a perfectly conducting surface, they are
solely determined by the dynamical dipole polarizability of
$\bar{H}$. Our treatment can be generalized to the case of finite
conductivity or non-zero temperatures by introducing corrections to
the Casimir-Polder interaction constants \cite{BKM}.  In particular,
the simple estimation of $\bar{H}$ quantum reflection constant $b$
in case of dielectric surface can be obtained using expression
(\ref{a4}) with the modified value of  $C_4$ coefficient:
\[
b \approx 4\sqrt{2mC_4\frac{\epsilon -1}{\epsilon +1}\digamma(\epsilon)}
\]
where $\epsilon$ is permittivity of the dielectric material. The
correction function $\digamma(\epsilon)$  calculated in \cite{DLP}
is approximately equal to $0.77$ in a wide range of $\epsilon$. The
estimated in such a way constant $b$ for polyethylene
($\epsilon\approx 2.3$) is $b\approx 1138.9$ a.u., which is
approximately twice smaller than in case of perfect conducting
surface. As a consequence the characteristic energy, for which
quantum reflection from polyethylene surface starts to be
significant ($kb\sim 1$, $E=1/(2mb^2)$) is four times greater, than
in case of perfect conducting surface.

In the following section we consider the confinement of
antihydrogen between material walls. In particular we show
how the properties of quantum reflection  are
related to the shifts and widths of the energy levels of
$\bar{H}$ atoms enclosed in cavities with conducting walls.

\section{Ultracold $\bar{H}$ between two conducting walls}

The partial reflection of ultra-cold $\bar{H}$ from the material
wall  enables the existence of quasi-stationary (decaying) states of
$\bar{H}$ in-between two  walls. In so far we will be interested in
the one-dimensional wave-function of such states.  We expect that
the distance $L$ between the conducting walls is large compared to
$|a_{CP}|$ so that the atom interacts with each of the walls
independently.  The boundary condition at the left wall ($z=0$) is
given by (\ref{IW}). To obtain the boundary condition at the right
wall one should replace in equation (\cite{IW}) $z$ by $L-z$.
 Far from the walls ($|a_{CP}|\ll z\ll L-|a_{CP}|$)
the wave function is

\begin{equation}\label{r}
\Phi(z)\sim \sin(kz+\delta_{CP})
\end{equation}
where $\delta_{CP}$ is the phase-shift produced by interaction with
the wall. Using the  symmetry of the problem one obtains the
quantization condition

\begin{equation}
kL+2\delta_{CP}=\pi n.
\end{equation}

We consider the low energy domain where that the scattering
length approximation is valid, so that $\delta_{CP}=-ka_{CP}$. Thus
we get for $k$

\begin{equation}
k=\pi n /(L-2a_{CP}),
\end{equation}

and so the energies of the box-like states are quantized.
The eigenvalues are modified by  quantum reflection according to

\begin{equation}
{\cal{E}}_n \equiv \varepsilon_n - i\frac{\Gamma_n}{2}=\frac{\pi^2
n^2}{2m (L-2 a_{CP})^2}\simeq \frac{\pi^2 n^2}{2m
L^2}(1+4\frac{\mathop{\rm Re} a_{CP}}{L})-i|\mathop{\rm Im}
a_{CP}|\frac{4\pi^2 n^2}{2m L^3}. \label{ebox}
\end{equation}

The widths of the states are given by

\begin{equation}
\Gamma_n=2b\frac{\pi^2 n^2}{2m L^3}=\frac{2\varepsilon_n^{(0)} b}{L}
\label{gbox}
\end{equation}

where $\epsilon_n^{(0)}=\pi^2n^2/(2mL^2)$ are the energy levels
unperturbed by the long range $\bar{H}$ -- wall interaction.

The level shifts and widths are determined by the value of
\emph{complex} scattering length ($b = 4 |Im a|$, c.f. eq.
\ref{aCP}) that characterizes quantum reflection. The lifetimes
increase with the wall separation as $L^3$ and decrease with the
excitation quantum number as $n^2$. These expressions are obtained
using the assumption that the scattering length approximation
(\ref{slapp}) is valid,  and restricts its validity for
$\varepsilon_n<10^{-11}$ a.u.

The solutions $\Phi_n(z)$,  corresponding to the complex energy
levels ${\cal{E}}_n$, are the eigen-functions of a non-self-adjoint
Hamiltonian. In fact, they are decaying quasi-bound states with
energy $\varepsilon_n-i\Gamma_n/2$. This implies that such states
obey the bi-orthogonality condition

\begin{equation}\label{biort}
\int_0^{\infty} \Phi_n(z)\Phi_k(z) dz =\delta_{nk}
\end{equation}
and differs from the "standard" expression by the absence of
complex conjugation of the $\langle {\rm bra}|$ function.

The expression for the width (\ref{gbox}) coincides with the
simple formula for the annihilation rate of the particles moving
freely  with the velocity $v=\sqrt{2\varepsilon^{(0)}/M}$
  between the walls separated by the
distance $L$

\begin{equation}
\Gamma=P(v)\omega,
\end{equation}
where $P(v)$ is the wall penetration probability and $\omega=v/L$ is
the frequency of "hits" on the wall. Substituting the probability of
annihilation on the wall $P(v)$ from (\ref{P}) we again obtain the
expression given in (\ref{gbox}), i.e.
$\Gamma=2\varepsilon^{(0)}\frac{b}{L}$. At high energies
($\varepsilon>10^{-8}$ a.u.) the annihilation probability
$P(v)\simeq 1$, and we get $\Gamma=\frac{v}{L} $. The lifetime of
such a "fast" particle is just its time of flight between the walls.

As an example, we take $L=10$ $\mu m$  which is much greater
than the length scale $b=0.115$ $\mu m$, that characterize the Casimir force.
 At
this wall separation distance, the ground state energy is
$\varepsilon_0=7.5$ $10^{-14}$ a.u. the width is $\Gamma_0=1.7$
$10^{-15}$ a.u. and corresponds to the lifetime $0.014$ s. We
compare this value to the time of flight between the walls
$t=L/v=0.0005$ s. Thus $\bar{H}$ in this  state bounces about 30
times before it annihilates. Since the life time is proportional to
$L^3$, and the passage time to $L^2$, the number of the bounces
between the walls grows linearly with $L$. Let us mention here that
the concept of quasi-stationary state itself is meaningful only in
case its life-time is much greater than the corresponding time of
flight between the walls.

\section{Antihydrogen in a wave-guide}
\label{wave-guide}
A laboratory demonstration of quantum reflection states could realized
by introducing $\bar{H}$ atoms through a slit between parallel conducting
walls separated by distance $L$. (The similar principle was used to
observe quantum motion of neutrons in the gravitational field of
Earth \cite{NPVnature,NPV}. Ignoring, for the moment, the influence
of gravitation on $\bar{H}$, we align the conducting
walls parallel to the gravitational field. The Schr\"{o}dinger
equation, that governs the motion of antiatoms inside the
wave-guide with transverse dimension $z$ and horizontal dimension
$x$, is

\begin{equation} \label {SchWG}
\left[ -\frac{\partial ^{2}}{2m\partial x^{2}}-\frac{\partial
^{2}}{2m\partial z^{2}}+V(z)+V(L-z)-E\right] \Psi (z,x)=0.
\end{equation}

We express the two-dimensional wave-function $\Psi (z,x)$ as a
series  of products of normalized transverse wave-functions
$\Phi_n(z)$ given in eq. (\ref{r}) and longitudinal plane-wave
functions $\exp(ip_n x)$:

\begin{equation}\label{Expand}
\Psi(z,x)= \sum_n C_n \exp(ip_n x)\Phi_n(z)
\end{equation}

where $C_n$ are the amplitudes of corresponding states, dependent on
the properties of the flux entering the wave-guide, and $p_n$ is
the horizontal momentum of the state with the transverse energy
$\varepsilon_n-i\Gamma_n/2$, so that

\begin{equation}\label{Phor}
p_n^2/2m+\varepsilon_n-i\Gamma_n/2=E.
\end{equation}

Let us mention that the eigen-functions $\Phi_n(z)$ of
not-self-adjoint Hamiltonian do not form a basis in the Hilbert
space \cite{Berg}. So far the expansion (\ref{Expand}) is an
approximation. We will show however that under certain conditions of
the wave-guide experiment few quasi-stationary states from
(\ref{Expand}) would be give exhaustive contribution  to the
measured flux, which verifies our approach.

 If the characteristic spatial dimension of
the incoming flux $H_0$ is much larger than the separation $L$ and
the distribution of the transverse velocity in the incoming flux
exceeds the value $1/L$, the first few  excited states are uniformly
distributed. Because the transverse energy
$\varepsilon_n-i\Gamma_n/2$ is complex, the horizontal momentum
$p_n$, for a given state, is also complex (c.f. \ref{Phor}).

The horizontal momentum at small transversal energy ($\varepsilon_n\ll E$)
is given by

\begin{equation}\label{phor}
p_n\simeq \sqrt{2mE} - {\sqrt{m/2E}}(\varepsilon_n + i{\Gamma_n/2})
= p -
  \frac{(\varepsilon_n-i\Gamma_n/2)}{v}
\end{equation}
where $p=\sqrt{2mE}$ and $v=p/m$. We can thus write for the
wave-function $\Psi(z,x)$

\begin{equation}
\Psi(z,x)= \exp(ipx)\sum_n C_n \exp(-\Gamma_n x/(2v))
\exp{-i\epsilon_n x/v} \Phi_n(z).
\end{equation}
Using the definition $t=x/v$ we relate the  decay  time   of
the transverse states $\Phi_n(z)\exp(-\Gamma_n t/2)$ to the
disappearance rate (along the $x$-axis) of the flux density inside
the wave-guide.

The integrated flux density $F=F_0\int|\Psi(z,d)|^2dz$
(where $F_0$ is a normalization constant) gives the number of counts
of $\bar{H}$ in the detector at the exit ($x=d$) of the wave-guide

\begin{eqnarray}
\label{flux}
\label{Fexit}
F &=& F_0\sum_n |C_n|^2 \langle\Phi_n |\Phi_n \rangle
\exp(-\Gamma_n \tau^{pass})
\\ \nonumber
&+& F_0\sum_{n\neq k}C_n^{*}C_k
\langle\Phi_n |\Phi_k \rangle
\exp(i(\varepsilon_k-\varepsilon_n)\tau^{pass})
\exp(-\frac{(\Gamma_n+\Gamma_k)\tau^{pass}}{2})
\end{eqnarray}
where we have introduced the passage time $\tau^{pass}=d/v$.

We denote the second term in this expression
as an "interference" term since its presence is due to
non-orthogonality of the decaying states,
$\langle\Psi_n |\Psi_k \rangle\neq \delta_{nk}$. The appearance of
such a term is not surprising. Indeed, decaying states are not
stationary and they don't have well defined energies and so they can be
thought of as a superposition of truly stationary states. Therefore
transitions between states with frequencies
$\omega_{nk}=\varepsilon_k-\varepsilon_n$ occur.
 The transitions manifest in the
appearance of the oscillating exponents
$\exp(i\omega_{nk}\tau^{pass})$ in the expression for the measured
flux (\ref{flux}). However, observation of such interference
requires rather good resolution in the horizontal velocities of the
initial flux. The uncertainty in passage time
$\delta\tau^{pass}=\tau^{pass}\delta v/v$ is of order of
$\omega_{nk}^{-1}$, and it follows that the limit for
horizontal velocity resolution is

\begin{equation}
\frac{\delta v}{v}<\tau^{pass}\omega_{nk}.
\end{equation}

For a broader distribution in the horizontal velocity, the
interference terms cancel due to fast oscillating exponents and
the expression for the flux takes the form

\begin{equation}
\label{Fexit1} F=F_0\sum_n \exp(-\Gamma_n \tau^{pass}).
\end{equation}

The time of flight $\tau^{pass}$ may be chosen in such a way that
only a few transverse states with the smallest
$\Gamma_n=n^2\Gamma_1$ contribute to the flux of $\bar{H}$ in the
wave-guide. This time of flight selects the states with small
transversal energy $\varepsilon_n$ and thus our expression for the
horizontal momentum (\ref{phor}) is justified. Let us also note that
such a choice of time of flight ensures that contribution of
quasi-stationary states to the flux at the exit of the wave-guide is
much greater than the contribution of all the "non-resonant" terms,
which were neglected in the approximative expansion (\ref{Expand})
of the total wave-function \cite{Baz}.  The typical length $d$ for a
flux of $\bar{H}$ with the horizontal velocity $v=10$ m/s should be
about 10 cm to ensure that only the ground transverse state passes
through the slit $L=10 \, \mu m$. Changing the slit separation leads
to an enhanced contribution from the excited transversal states, due
to the dependence (\ref{gbox}) of the decaying rates $\Gamma_n$ on
$L$. The measured function $F(L)$ provides information on the
parameter $b$. In particular, if only the ground state passes
through the guide, one gets the following ($C_1$ and $F_0$
independent) expression for $b$

\begin{equation}
b=\left(\frac{\partial \ln F(L)}{ \partial L}\right)\frac{mL^4 }{
3\pi^2\tau^{pass}}.
\end{equation}

We stress  that the quantity $b$ above is very sensitive to
the Casimir force since
quantum reflection in the guide selects only those
antiatoms that have interacted with the surface through
the van der Waals - Casimir potential  (all others are annihilated).

\section{Gravitational effects}

We have shown that quantum reflection takes place for very slow
anti-atoms. This fact could be exploited to study the gravitational
properties of anti-atoms. In this section we discuss the possibility
for measuring the gravitational force acting on antihydrogen.
 We distinguish between the gravitational mass, which
  we refer to as $M$ and inertial mass, hereafter denoted by
  $m$.
We consider an $\bar{H}$ atom bouncing above a conducting
surface in the gravitational field of Earth. Confinement is
achieved by quantum scattering by the van der Waals- Casimir
potential from below, and by the gravitational field from above. The
characteristic length and energy scales are
\begin{eqnarray}\label{scaleL}
l_0 &=&\sqrt[3]{\frac{\hbar^{2}}{2mMg}}, \\
\varepsilon &=&\sqrt[3]{\frac{\hbar^2M^2g^2}{2m}}.\label{scaleE}
\end{eqnarray}

  The characteristic scale of the gravitational energy of $\bar{H}$
  is $\varepsilon=2.211$ $10^{-14}$
  a.u., whereas the corresponding length scale is $l_0 =5.871$
$\mu m$ and is much greater than the length scale, ($b=0.115$ $\mu m$ ), of the Casimir force.
 Thus, the gravitational field does not
significantly affect the atom-wall interaction. The energy levels
and the corresponding wave-functions are

%

\begin{eqnarray}
E_n&=&\varepsilon \lambda_n, \\
\Phi_n(z)&=&C \mathop{\rm Ai}(z/l_0-\lambda_n)
\end{eqnarray}

where $\mathop{\rm Ai}(x)$ is the Airy function \cite{Ab} and the
eigenvalues $\lambda_n$ are found by enforcing the condition $\mathop{\rm
Ai}(-\lambda_n)=0 $. The modification in the values for the eigenvalues $\lambda_n$
due to  quantum reflection are obtained by
matching the wave-function of $\bar{H}$ reflected from the wall
(which at the distance $|a_{CP}|\ll z\ll l_0$ has the asymptotic
form $\Phi(z)\sim 1-z/a_{CP}$) to the gravitational wave-function
$\mathop{\rm Ai}(z/l_0-\widetilde{\lambda})$, where
$\widetilde{\lambda}$ is a modified eigenvalue. Taking into account
that in the matching region $z/l_0\ll 1$, we obtain the following
equation for $\widetilde{\lambda}$

\begin{equation} \label{lambda}
\frac{\mathop{\rm Ai}(-\widetilde{\lambda})}{\mathop{\rm
Ai}'(-\widetilde{\lambda})}=-a_{CP}/l_0.
\end{equation}

Because $a_{CP}/l_0\ll1$,  $\widetilde{\lambda}$ can be expressed
as a perturbation series in terms of the gravitational eigenvalues
$\lambda_n$

\begin{equation}
\widetilde{\lambda}_n=\lambda_n+\delta\lambda_n
\end{equation}
with $\delta \lambda_n \ll 1$. This leads an expression
for the modified eigenvalues $\widetilde{\lambda}_n$

\begin{equation}\label{lambnew}
\widetilde{\lambda}_n=\lambda_n+a_{CP}/l_0.
\end{equation}

 From it, we obtain modified energy levels and the
corresponding widths

\begin{eqnarray}
E_n&=&\varepsilon (\lambda_n + \frac{\mathop{\rm Re}
a_{CP}}{l_0})\label{Egrav}, \\
\Gamma_n&=& \varepsilon \frac{b}{2l_0}\label{Wgrav}
\end{eqnarray}

The widths of the gravitational states
(\ref{Wgrav}) are proportional to the combination
$\varepsilon/l_0$.  Using (\ref{scaleL}) and (\ref{scaleE}) we find
that this ratio is equal to the gravitational force
$\varepsilon/l_0=Mg$ so that

\begin{equation}
\Gamma_n = \frac{b}{2} Mg.
\end{equation}

The value of the ground state energy of $\bar{H}$ in the
gravitational field of Earth turns out to be $E_1=5.17$ $10^{-14}$
a.u. The corresponding life time is

\begin{equation}
\tau=\frac{2\hbar}{Mgb} \simeq 0.1 \mbox{s}. \label{time}
\end{equation}

We note factorization of gravitational effects (appearing in the
above formula via  factor $Mg$) with the effect of quantum
reflection, appearing through the constant $b$. Such a factorization
is a consequence of the small ratio in the characteristic scale
$b/l_0\simeq 0.02$.

We  find that the widths of the gravitational states
(\ref{Wgrav}) are independent of the energy (for the energies
$E_n<10^{-11}$ a.u.) to  the first order in the small ratio
$b/\lambda_n$. This is understood using the following
simple argument. The frequency of the atom bouncing above the
surface in the gravitational field is $\omega \sim 1/\sqrt{E}$,
whereas the probability (\ref{P}) of  annihilation on the wall
is $P\sim \sqrt{E}$. Combining the two, we obtain an energy
independent expression for the width,  $\Gamma=\omega
  P$. That is, all $\bar{H}$ atoms with $E_n<10^{-11}$ a.u. bouncing
on the surface
  survive
for $0.1$ s.

   The measurement of this lifetime can be achieved by
   monitoring the spatial decay
of the flux of antiatoms
   with the velocity $v$ parallel to the surface.
    Such a flux, as function of the distance $d$
    and integrated over vertical dimension $z$ is

\begin{equation}
F(d)=F_0 \exp(-\frac{\varepsilon}{\hbar} \frac{b}{2l_0} \tau^{pass})
=\exp(-\frac{Mgb}{2\hbar}\tau^{pass} ).
\end{equation}

If one shifts the position $d$ of the detector and measures the
spatial decay $F(d)$, one can infer the gravitational force $ Mg$ acting
on the antiatom since

\begin{equation}
Mg=-2\frac{\partial \ln F(d)}{\partial d} \frac{\hbar v}{b}. \label{mg}
\end{equation}

\section{Surface states for atoms and antiatoms}

Our treatment of quantum reflection of antiatoms makes use of the
fact that the antiatom-wall interaction is strongly absorptive at short
distance, and is mathematically expressed by
absorption condition (\ref{fullabs}). In this section we  study
the case where absorption on the surface is not necessarily strong.
Such a situation may correspond to the partial loss of ordinary
atoms that undergo inelastic interaction at the surface, or to
  reflection of antiatoms from partially transparent
evanescent-wave atomic mirror.

 Below we outline a resonance-type dependence of the atom loss
probability on the properties of the short range interaction in case
of the weak surface absorption. We show that such  resonance
behavior stems from the presence
  of narrow near-threshold atom-surface states
\cite{States}.

For the purpose of illustration, and in order to allow
 an analytic description, we replace the exact van der Waals-Casimir potential
by the model potential (\ref{Vm}). We do not impose the full absorption boundary condition,
 instead
the \emph{complex} phase-shift $\delta_s$, produced by the
short-range interaction is used.

For the model potential $V_m$, the analytic expression for the scattering length
(\ref{amex})
 simplifies  substantially  since $\xi=2C_3\sqrt{2m/C_4}\gg 1$.
and using the asymptotic representation of the Hankel functions
$H_1^{(1,2)}(\xi)\sim \sqrt{2/(\pi \xi)} \exp(\pm i(\xi-3\pi/4)$ we
obtain
\begin{equation}\label{amas}
a_m=-\frac{l}{2}\xi\cot(\xi-\delta_1-i\delta_2-3\pi/4).
\end{equation}

We explore the dependence of $a_m$ on the real part
of the short-range phase-shift $\delta_1$.
  We find that  $a_m$ is
  an oscillating function of $\delta_1$.
  The smaller the imaginary part
$\delta_2$, the more prominent are the oscillations. The maxima occur
when
\begin{equation}\label{res}
\delta_1^{res}=\xi-3\pi/4-\pi n \mbox{, } n=1,2,...
\end{equation}
This expression is simply the condition for
the appearance of a  new state in the  potential well of an
attractive van der Waals-- Casimir potential and a short-range
core.

The existence of the narrow near-threshold state leads to a
probability loss
  $P=4k|\mathop{\rm Im} a_m|$ that varies rapidly with variations
in the phase $\delta_1$. At its maximum, the probability
loss (for the above mentioned case when $\xi\gg1$ and
$\delta_2\ll1$) is

\begin{equation}
P_{max}=2kl\frac{\xi}{\delta_2}
\end{equation}
and the width of such a maximum is equal to $\delta_2$.

This  example illustrates the influence of  near-threshold
singularities on the S-matrix and scattering observables. To find
the position of the S-matrix poles in the complex $k$-plane, we
express the S-matrix in the vicinity of its
pole

\begin{equation}
S=\frac{z+k}{z-k}.
\end{equation}

Taking into account that for small $k\ll |z|$ we have $S=1-2ik
a$, we get for the \emph{near-threshold} pole position

\begin{equation}
z=i/a.
\end{equation}

Using a Taylor expansion of the scattering length $a_m$
(\ref{amas}) in the vicinity of the resonance $\delta_1^{res}$,  the
position of the near-threshold pole can be written as

\begin{equation}
z\simeq \frac{2i}{l\xi}\Delta \delta_1-\frac{2\delta_2}{l\xi}
\end{equation}

where $\Delta \delta_1=\delta_1-\delta_1^{res}$ is the deviation of
the short-range phase-shift from the
  resonance value.

  The maximum
in the loss probability occurs when the  pole $z$ of the S-matrix
crosses the real axis of the complex $k$-plane. In the absence of
absorption, the pole lies on the imaginary axis of the complex
$k$-plane and when it crosses the real axis the
scattering length approaches positive infinity. In the presence of absorption
the pole has a real negative component. The smaller the shift from the imaginary
axis, the more pronounced the effect.

  As an illustration, we consider
the values $\delta_1$ and $\delta_2$ obtained within the  model potential

\begin{equation}
V_s(z)= \left\{
\begin{array}{cll}
U\exp(-i\varphi)\exp\left((z_0-z)/\rho\right) & \mbox{if} & z \geq z_0 \\
U\exp(-i\varphi) & \mbox{if} & z < z_0
\end{array}
\right.
\end{equation}

Here the argument of the complex exponent $\varphi$
relates the real and imaginary part of the complex
potential and $\rho$ is the diffuseness radius.  This
model allows analytic solutions and facilitates our
understanding of inelastic processes
nearby and inside the wall.

For a strong complex
exponential potential $V_s$ where $|\mathop{\rm
Im}\rho\sqrt{2MU}|\gg 1$,  the scattering length is \cite{KPV}
\begin{equation}
a_{s}\simeq z_0+\rho\left( 2\gamma+2\ln(\rho\sqrt{2m|U|})-i\varphi
\right)
\end{equation}

where $\gamma\simeq 0.577$ is the Euler constant.

We first consider the case  where the diffuse radius $\rho$ of the
exponential potential is much smaller than the local de Broglie
wavelength $\lambda_{B}(z_0)=1/\sqrt{2mC_3/z_0^3}$ at the distance
$z_0$. Here the short range potential $V_s$ varies rapidly when
compared to variations of the de Broglie wavelength. At distances
$\rho\ll (z-z_0) \ll \lambda_{B}(z_0)$ the wave-function is

\begin{equation}
\Psi_{s}=1-\frac{z-z_0}{a_{s}-z_0}.
\end{equation}

Matching it with wave-function (\ref{Phim}) and taking into
account the asymptotic properties of the Hankel function with a
large argument

\begin{equation}
\xi_0\equiv 2\sqrt{2mC_3(1/z_0+C_3/C_4)})\gg 1
\end{equation}

we obtain

\begin{eqnarray}
\label{deltas} \delta_1&\simeq&\xi_0-3\pi/4\\
  \delta_2 &=& \varphi\rho
/\lambda_{B}(z_0)=\varphi\rho\sqrt{2MC_3/z_0^3}\label{delta2}.
\end{eqnarray}

where we have kept only the leading terms in
$\rho/\lambda_{B}(z_0)$.

Thus, the  limit $\rho/\lambda_{B}(z_0)\rightarrow 0$
implies vanishing absorption ($\delta_2 \rightarrow 0$) and
corresponds to the case where the  reflecting wall is placed at position
$z_0$.


The weak absorption limit $\delta_2\ll1$ can be realized
by an
 optical evanescent wave that generates an effective
repulsive potential. The detailed study of this case is beyond the
scope of our paper; for the purpose of our qualitative analysis it
is only important that the imaginary part of the phase-shift
$\delta_2$ turns to be small when  $\rho/\lambda_{B}(z_0) \ll 1$.
  At the same time the phase-shift
$\delta_1$ is well approximated by expression
(\ref{deltas}).


In the case  $\rho/\lambda_{B}(z_0)\geq 1$, the
short range potential varies slightly with respect to
the scale of change of
$\lambda_B$ and one can use the semiclassical expression for the
phase-shift

\begin{equation}
\delta_2=\mathop{\rm
Im}\int_{-\infty}^{\infty}\sqrt{-2m(V_s(z)+V_m(z))}dz
\end{equation}

Thus  a  shallow but extended imaginary potential can
produce a large imaginary phase-shift $\delta_2\gg 1$. A
a necessary condition for the validity of the strong
absorption boundary condition is that  the  short-range interaction
varies slowly in comparison with the variation in the de Broglie wavelength
$\lambda_B(z_s)$ at $z_s$, the distance where the short-range
interaction is important.  For $\bar{H}$ the
characteristic de Broglie wavelength at $z_s\simeq 1$
a.u. is $\lambda_B(z_s)\sim 0.05$ a.u. and the scale
$\rho$ of the short-range inelastic interactions is about $1$ a.u.
 Well within the range where the strong absorption boundary condition is
 valid.


\begin{figure}
  \centering
\includegraphics[width=120mm]{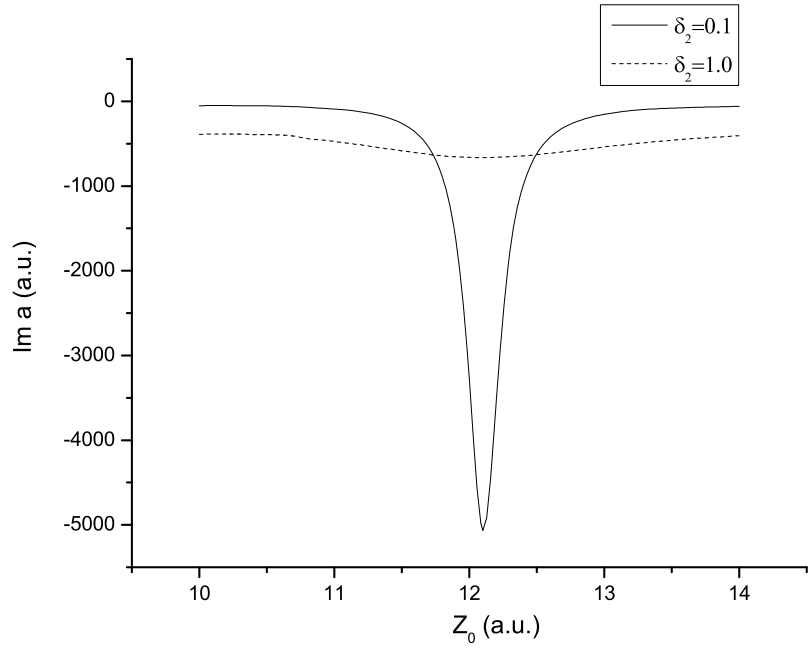}
  \caption{Imaginary part of the scattering length as a function of the position
of the "reflecting wall".}\label{Fig3}
\end{figure}



\begin{figure}
  \centering
\includegraphics[width=120mm]{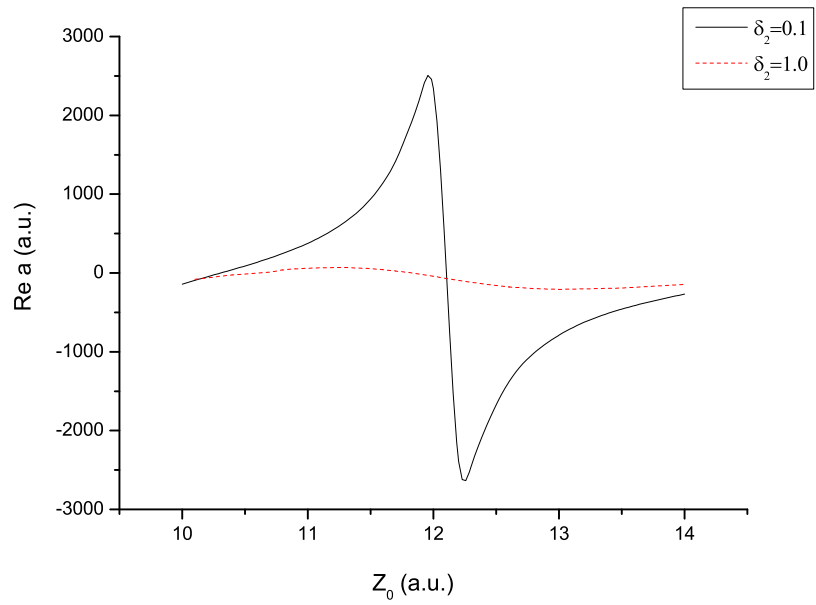}
  \caption{Real part of the scattering length as a function of the position of
the "reflecting wall". }\label{Fig4}
\end{figure}


We substitute
(\ref{deltas}) into (\ref{amas}) to get

\begin{equation}\label{amwall}
a_m=-\frac{l}{2}\xi \cot(\xi-\xi_0-i\delta_2)
\end{equation}

where $n=(\xi_0-\xi)/\pi$ is the number of
atom-surface bound states.

The scattering length $a_m$ exhibits resonance behavior when the
position of the "reflecting wall" is chosen in such a way that
a new state appears at threshold, i.e. $\xi_0(z_0)-\xi\rightarrow
\pi n_0$. In the vicinity of $z_0$ it is
useful to represent the scattering length (\ref{amwall}) in terms of
a  Breit-Wigner form

\begin{eqnarray}\label{BWam}
\mathop{\rm
Im}a_m=-\frac{l\xi}{2}\frac{\delta_2}{(\xi-\xi_0-\pi n_0)^2+\delta_2^2}, \\
\mathop{\rm Re}a_m=-\frac{l\xi}{2}\frac{\xi-\xi_0-\pi
n_0}{(\xi-\xi_0-\pi n_0)^2+\delta_2^2}.
\end{eqnarray}

In Fig. \ref{Fig3} we plot the imaginary part of $a_m$ as a function of
$z_0$ for two values of the imaginary phase-shift $\delta_2=0.1$ and
$\delta_2=1.0$. We vary $z_0$ from $10$ a.u. to $15$ a.u. Fig.
\ref{Fig4} illustrates the behavior of the real part of the
scattering length for the two values of $\delta_2$ given above.
We find a resonance-like increase in atom loss when the
short range potential is centered around $z_0=12.1$ a.u. This
resonance is very prominent for weak absorption when $\delta_2=0.1$,
and it is practically washed out in case of stronger absorption
$\delta_2=1.0$.

In the hypothetical case where the short range part of
the (anti)atom-wall interaction can be varied in a controlled way
(e.g. by reflection from the optical evanescent wave) one could
accurately measure the properties of the (anti)atom-wall potential
by careful tuning of resonances and
 fitting the experimental data to eq. (\ref{BWam}). In this way, one could
measure the values of the  $C_3$ and $C_4$ coefficients.

In conclusion, we find that the main difference in
quantum reflection in cases of strong and weak absorption by the
material wall is that in the latter case, narrow atom-surface
states can be formed. The position of these states depends on the
short range interaction. The presence of a near-threshold state
strongly enhances the probability of atom loss, and might  be used
for resonant spectroscopy (e.g. with the use of
optical evanescent wave). In case of strong absorption, the
near-threshold states are completely destroyed by annihilation.

\section{Conclusions}

We presented a theoretical account of
cold antiatoms interacting with a material wall. We focused on the
antihydrogen - wall system where the latter is represented by
an ideally conducting surface. We have
shown that, as  antiatoms approach the surface in an
attractive potential they are not necessarily annihilated.
If sufficiently cold, they may be totally reflected. Reflection occurs
in the region where the long range tail of the van der Waals - Casimir
interaction dominates. We calculated the reflection probability
as function of the incident energy of antiatoms. We analyzed
the behavior of the reflected wave as a function of distance
from the surface and introduced the notion of a "reflection distance".
We have also shown that quantum reflection
allows the existence of long-lived metastable states of
$\bar{H}$ in the conducting cavity.

We pointed out that scattering of ultra-cold $\bar{H}$ in an
attractive van der Waals - Casimir potential is sensitive to
retardation in the antiatom-wall
interaction. We explored the possibility of performing experimental
measurements with
$\bar{H}$ by introducing it into  a wave-guide. The rate of
the $\bar{H}$ annihilation on the wall of the guide could  provide important
information on retardation effects in matter-antimatter systems.
In the strong absorption limit, such
measurements are not "contaminated"
by effects produced on the wall and which are characterized by poorly known
short-range parameters. However, they could be employed to study
the asymptotic properties of $\bar{H}$-
wall interaction.

We also proposed measurement of effects due to
gravity  acting on the antihydrogen. In particular
we found that the life-time of $\bar{H}$,
in the combined gravitational and the mirror image potential of a conducting plate,
to be on the order of $0.1$ s. The established dependence of
that lifetime,  $\tau=2\hbar/ (Mgb)$,  on the gravitational force allows
the possibility for measuring the  gravitational properties of antimatter.

Finally, we outlined the role of the near-threshold atom-surface
states the  where absorption on the surface is small.
For antiatoms, weak absorption could arise with the introduction
of an evanescent wave that prevents antiatoms from direct contact
with the wall. We showed that a resonant increase in the absorption
rate can be observed if the atom-wall interaction is "tuned" to form
an atom-wall state at the threshold. We note the possibility of
exploiting resonant absorption for studies of atom-surface
interactions.

\section{Acknowledgment}
The research was performed under support from the Wenner-Gren
Foundations, the Swedish Natural Research Council and the Russian
Foundation for Basic Research grant 02-02-16809.

\bibliographystyle{unsrt}

\bibliography{hbarwall-bzedit1}

\begin{thebibliography}{10}

\bibitem{LJ}
J.E. Lennard-Jones.
\newblock {\em Trans. Faraday Soc.}, 28:333, 1932.

\bibitem{yu93}
I.A. Yu, J.M. Doyle, J.C. Sandberg, C.L. Cesar, D.~Kleppner, and T.J. Greytak.
\newblock {\em Phys. Rev. Lett.}, 71:1589, 1993.

\bibitem{ExCh1}
F.~Shimizo.
\newblock {\em Phys. Rev. Lett.}, 86:987, 2001.

\bibitem{ExCh0}
T.~A. Pasquini, Y.~Shin, C.~Sanner, M.~Saba, A.~Schirotzek, D.~E. Pritchard,
  and W.~Ketterle.
\newblock {\em Phys.Rev.Lett}, 93:160406, 2004.

\bibitem{CP}
H.B. Casimir and D.Polder.
\newblock {\em Phys. Rev.}, 73:360, 1948.

\bibitem{DLP}
I.E.Dzyaloshinskii, E.M. Lifshitz, and L.P. Pitaevskii.
\newblock {\em Adv. Phys.}, 10:165, 1960.

\bibitem{Bord}
M.~Bordag, U.~Mohideen, and V.M. Mostepanenko.
\newblock {\em Phys. Rep.}, 353:1, 2001.

\bibitem{FJM}
H.~Friedrich, G.~Jacoby, and C.G. Meister.
\newblock {\em Phys. Rev. A}, 65:032902, 2002.

\bibitem{ExCh2}
A.~Landragin et~al.
\newblock {\em Phys. Rev. Lett.}, 77:1464, 1996.

\bibitem{ExCh3}
C.I.~Sukenik et~al.
\newblock {\em Phys. Rev. Lett.}, 70:560, 1993.

\bibitem{ExCh4}
A.Shih and V.A. Parsegian.
\newblock {\em Phys. Rev. A}, 12:835, 1975.

\bibitem{AH1}
H.~Amoretti et~al.
\newblock {\em Nature (London)}, 419:456, 2002.

\bibitem{AH2}
G.~Gabrielse et~al.
\newblock {\em Phys. Rev. Lett.}, 89:213401, 2002.

\bibitem{frie04a}
H.~Friedrich and J.~Trost.
\newblock {\em Physics Reports}, 397:359, 2004.

\bibitem{juri04}
A.~Jurisch and h.~Friedrich.
\newblock {\em Phys. Rev. A}, 70:032711, 2004.

\bibitem{frie04}
H.~Friedrich and A.~Jurisch.
\newblock {\em Phys. Rev. Letters}, 92:103202.

\bibitem{DMB}
M.Marinescu, A.~Dalgarno, and J.F. Babb.
\newblock {\em Phys. Rev. A}, 55:1530, 1997.

\bibitem{BKM}
J.F. Babb, G.L. Klimchitskaya, and V.M. Mostepanenko.
\newblock {\em Phys. Rev. A}, 70:042901, 2004.

\bibitem{W}
G.N. Watson.
\newblock {\em A treatise on the theory of Bessel functions}.
\newblock Cambridge University Press, 1922.

\bibitem{MM}
M.F. Mott and H.S.W. Massey.
\newblock {\em The theory of atomic collisions}.
\newblock Oxford, Clarendon Press, 1965.

\bibitem{AV}
A.~Yu. Voronin.
\newblock {\em Phys. Rev. A}, 67:062706, 2003.

\bibitem{Babikov}
V.~V. Babikov.
\newblock {\em {The method of phase functions in quantum mechanics}}.
\newblock Moscow: Nauka (in Russian), 1967.

\bibitem{Calog}
F.~Calogero.
\newblock {\em Phase Approach to Potential Scattering}.
\newblock New-York: Academic, 1967.

\bibitem{Pokrovskii1}
S.~K.~Savvinykh V.~L.~Prokovskii and F.~K. Ulinich.
\newblock {\em Sov.Phys. JETP}, 34:879, 1958.

\bibitem{Pokrovskii2}
S.~K.~Savvinykh V.~L.~Prokovskii and F.~K. Ulinich.
\newblock {\em Sov.Phys. JETP 34}, 34:1119, 1958.

\bibitem{Phonon1}
A.~Siber B.~Gumhalter and J.P. Toennies.
\newblock {\em Phys. Rev. Lett.}, 83:1375, 1999.

\bibitem{Phonon2}
B.~Gumhalter A.~Siber.
\newblock {\em Phys. Rev. Lett.}, 90:126103, 2003.

\bibitem{NPVnature}
V.V.~Nesvizhevsky et. al.
\newblock {\em Nature}, 415:297, 2002.

\bibitem{NPV}
V.V.~Nesvizhevsky et~al.
\newblock {\em Phys. Rev. D}, 67:102002, 2003.

\bibitem{Berg}
T.~Berggren.
\newblock {\em Nucl. Phys.}, page 265, 1968.

\bibitem{Baz}
Ya. B.~Zeldovich A.I.~Baz and A.M. Perelomov.
\newblock {\em Scattering, Reactions and Decays in the Nonrelativistic Quantum
  Mechanics}.
\newblock Moscow, Nauka (in Russian), 1966.

\bibitem{Ab}
M.~Abramowitz and I.E. Stegun.
\newblock {\em {\it Handbook of mathematical Functions}}.
\newblock Dover Publ., New York, 1965.

\bibitem{States}
E.G. Lima, M.Chevrollier, O.Di Lorenzo, P.C. Segundo, and M.~Oria.
\newblock {\em Phys. Rev. A}, 62:013410, 2000.

\bibitem{KPV}
V.A. Karmanov, K.V. Protasov, and A.~Yu. Voronin.
\newblock {\em Eur. Phys. J.}, 8:A 429, 2000.

\end{thebibliography}

\end{document}